\documentstyle[sprocl]{article}

\input{psfig}

\def\Journal#1#2#3#4{{#1} {\bf #2}, #3 (#4)}

\def\NPB{{\em Nucl. Phys.} B}
\def\PLB{{\em Phys. Lett.}  B}
\def\PRL{\em Phys. Rev. Lett.}
\def\PRD{{\em Phys. Rev.} D}

\def\be{\begin{equation}}
\def\ee{\end{equation}}
\def\bea{\begin{eqnarray}}
\def\eea{\end{eqnarray}}

\begin{document}

\title{ VORTEX PERCOLATION AND CONFINEMENT } 

\author{K. LANGFELD }

\address{Institut f\"ur Theor. Physik, Universit\"at T\"ubingen, 
Auf der Morgenstelle 14 \\ D-72076 T\"ubingen, Germany 
\\E-mail: langfeld@alpha8.tphys.physik.uni-tuebingen.de }


\maketitle\abstracts{ The vortex state which arises from a 
projection of SU(2) to $Z_2$ gauge theory is studied at finite 
temperatures with a special emphasis on the deconfinement phase 
transition. }

\section{Introduction}

Since the time that the universality of the parton model, designed for 
describing deep inelastic nucleon scattering processes,  
was understood by means of perturbative QCD, the non-liberation 
of the partons, i.e. confinement, even at high energies is still seeking 
an explanation. Subsequently, it was observed that a static quark anti-quark 
potential which linearly rises at large distances successfully describes 
the charmonium spectra. With the advent of the numerical 
approach to lattice gauge theories, the assumption of a linear confining 
potential was put onto solid grounds. The role of these 
simulations was not only to establish quark confinement from first 
principles Yang-Mills theory, but also to reveal the origin of the 
linear rise: the color-electric flux of the quark anti-quark pair 
is squeezed into a flux tube. In the last decade of intense 
investigations of Yang-Mills theories, it has turned out crucial 
for {\tt understanding confinement} to reveal the so-far hidden 
roots of the color-electric flux tube formation. A substantial progress 
in this direction was made by realizing that {\it projection techniques} 
are a convenient tool for theses purposes: the link variables which 
represent an actual configuration of the SU(N) lattice gauge theory are 
projected onto elements of a group ${\cal G}$ the number of degrees of 
freedom of which counts less than that of the SU(N) theory. 
The choice of ${\cal G}$ is constrained by demanding that the 
string tension is (almost) unchanged by projection. Preserving 
confinement while reducing the number of degrees of freedom stirs 
the hope for clearing up those degrees relevant for confinement. 
Choosing ${\cal G} = U(1) _{compact}$ uncovers color-magnetic monopoles 
and a (color) photon~\cite{tho81}. In this setting, 
numerical evidence is found that color electric flux tube formation 
is due to a dual Meissner effect which is generated by the condensation 
of the color magnetic monopoles~\cite{bali}. Here, we will not 
follow these lines. We will study the case of a SU(2) gauge theory and 
will reduce the number of degrees of freedoms by employing 
${\cal G} = Z_2$~\cite{deb96}, \cite{deb97}. This case reveals 
$Z_2$ vortices as the only configurations of the projected theory. 
I will discuss the role of the vortices for confinement at zero as well 
as finite temperature, and will address the drastic change of the 
vortex properties at the deconfinement phase transition.

\section{ The SU(2) vortex vacuum }\label{sec:2} 

\subsection{ Construction of the theory of vortices } 

The so-called center projection of SU(2) gauge theory on a $Z_2$ 
gauge theory is accomplished by successive steps of gauge fixing 
and projection. In a first step one exploits the gauge degrees 
of freedom for minimizing the off-diagonal elements of the 
SU(2) link variables $U_\mu (x)=:U(X)$, i.e. 
\be 
U(X) \; = \; \left( \matrix{ a_0 + i a_3 & a_2 +i a_1 \cr 
-a_2 + i a_1 & a_0 - ia_3 \cr } \right) \; , 
\hbox to 1cm {\hfill } 
\sum_X \left( a_1^2(X) + a_2^2(X) \right) \rightarrow 
\hbox{min} \; , 
\label{eq:1} 
\ee 
where the constraint $a_0^2(X) +a_1^2(X) +a_2^2(X)+a_3^2(X)=1$ must be 
obeyed for all $X$. {\it Abelian projection } $U (X) \rightarrow 
\bar{U}(X) \in U(1) $ is defined by setting the off-diagonal elements 
to zero and normalize $\bar{U}$ to ensure $\bar{U} \bar{U}^\dagger =1$. 
After this first step of projection the reduced theory is a compact U(1) 
gauge theory with (color) photons and magnetic monopoles as degrees 
of freedom. In a second step, one exploits the residual U(1) gauge 
degree of freedom for a minimization of $\sum _X a_3^2(x)$. {\it 
Center projection} $\bar{U}(X) \rightarrow U^\prime(X) \in Z_2$ 
is performed by setting $a_3(X)=0$ and by normalizing $U^\prime 
U^{\prime \, \dagger }=1$. After this second step the link elements 
of full SU(2) gauge theory are reduced to elements $\{\pm 1\}$, i.e. 
$U \in SU(2) \rightarrow U^\prime \in Z_2$. The reduced theory possesses 
a residual $Z_2$ gauge degree of freedom. 

\begin{figure}[t]
\rule{5cm}{0.2mm}\hfill\rule{5cm}{0.2mm}
\centerline{
\hspace{-.6cm} 
\psfig{figure=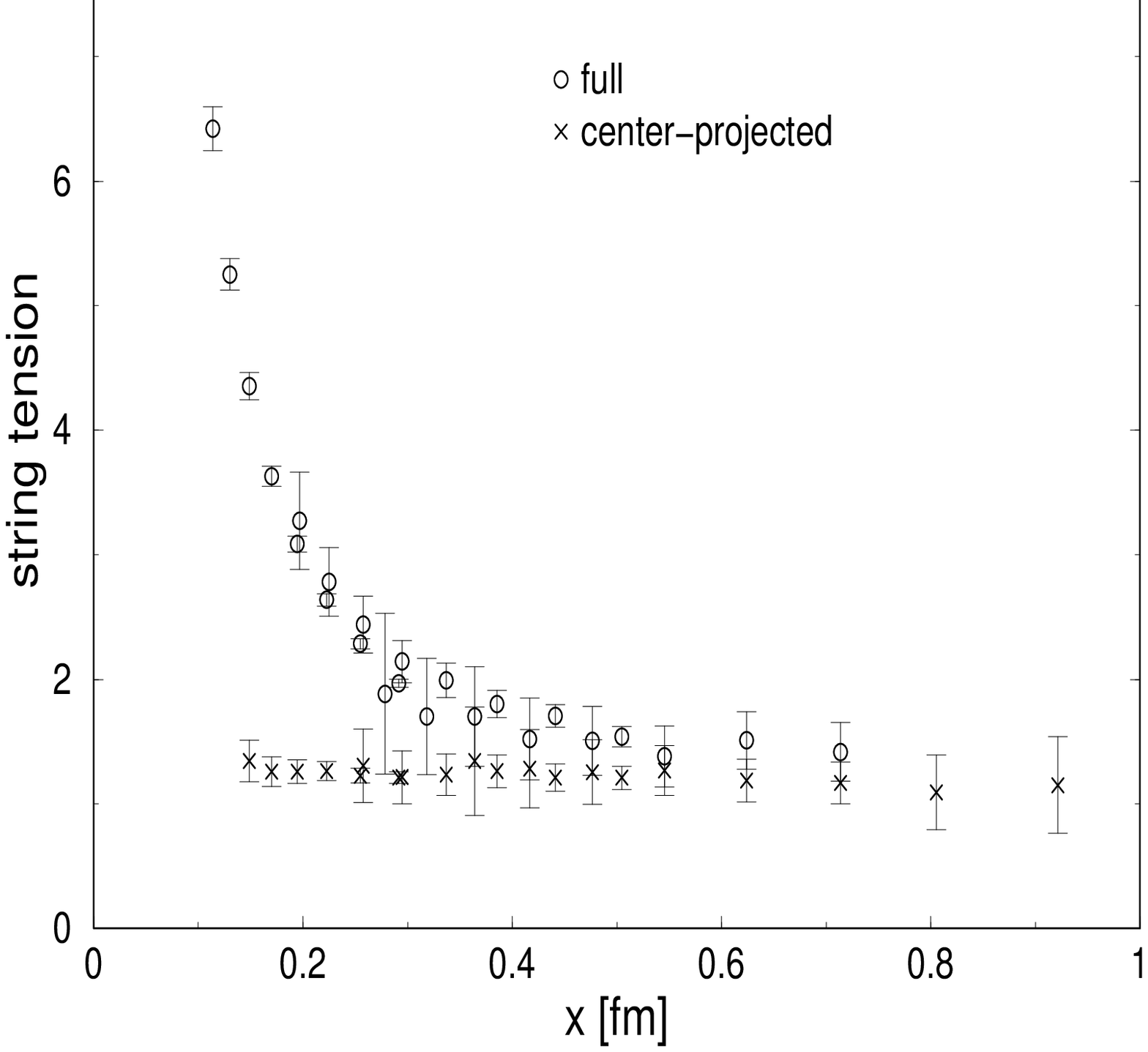,height=1.5in}
\hspace{1cm} 
\psfig{figure=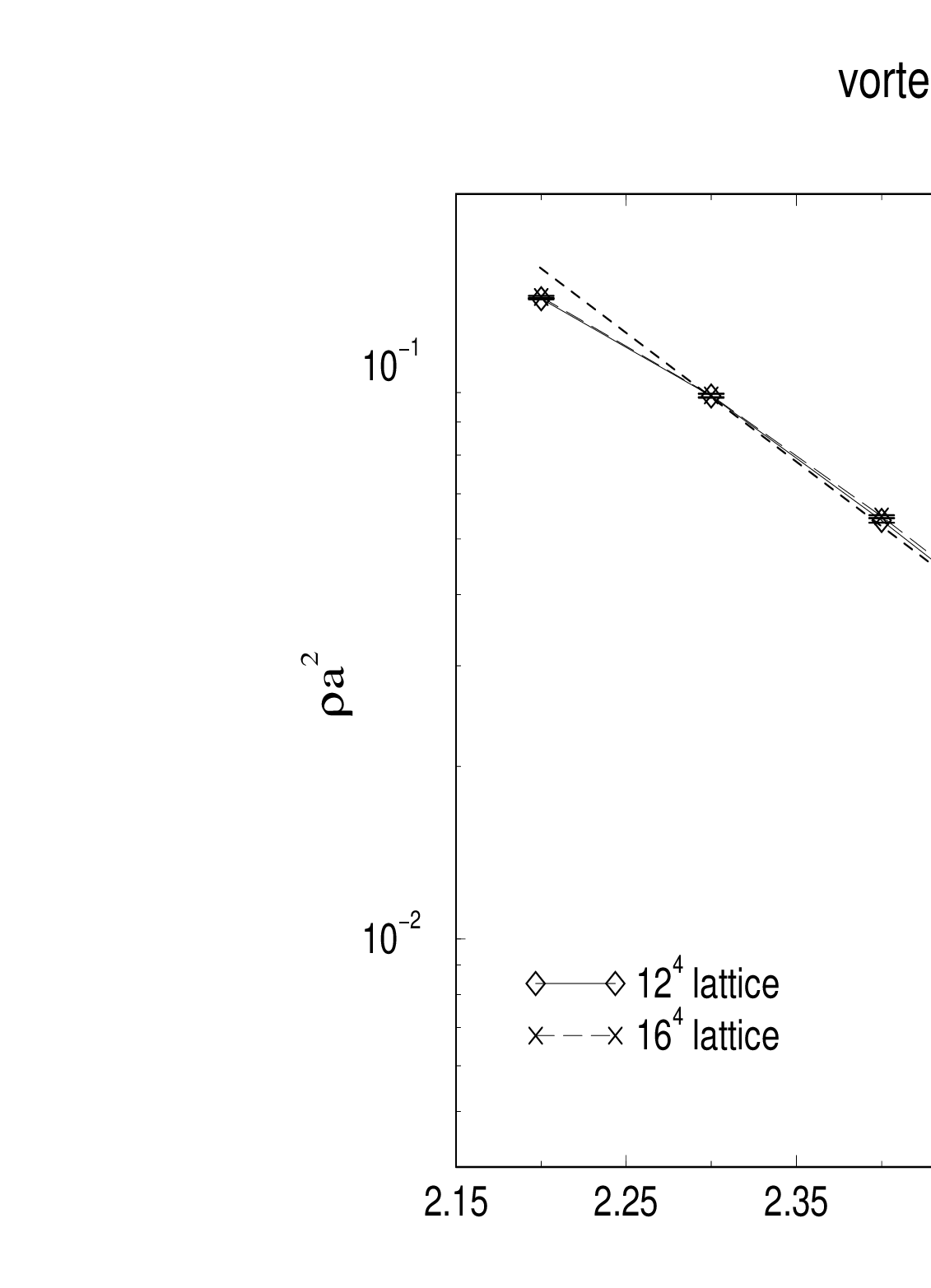,height=1.5in}
}
\rule{5cm}{0.2mm}\hfill\rule{5cm}{0.2mm}
\caption{ The Creutz ratios $\chi (r,r)$ as function of $r$ in physical 
  units (left panel); the vortex area density in units of the lattice 
  spacing $a$ as function of $\beta $ (right panel).   \label{fig:1}}
\end{figure}
A $Z_2$ gauge theory can be phrased as a theory of vortices. For this 
purpose, we choose a 3-dimensional hypercube of space-time. We 
define that a vortex pierces a plaquette $p$ if the product of the 
center projected links which are part of $p$ yield $-1$, i.e. 
$v_p(x) := \prod _{l \in p} U^\prime _l(x) = -1 $. The plaquette 
which is pierced by a vortex is part of two hypercubes, and we 
consider the $Z_2$ flux line which connects the center of the 
corresponding hypercubes as part of the vortex. By construction, the 
vortex lines fall on top of the links of the 3-dimensional dual lattice. 
For revealing the string type nature of these vortex links. we introduce 
$n$ as the number of vortices which pierce the plaquettes $p$ of 3-dimensional 
hypercube $c$. Resorting to the identity 
$1 = \prod _{p \in c } v_p(x) = (-1)^n $ 
one finds that the number $n$ is even. A vortex never ends inside a cube 
implying that the vortices form closed lines within the 3-dimensional 
hypercube. Since one is free in choosing the 3-dimensional hypercube 
out of the 4-dimensional space time, we conclude that the vortices from 
closed world sheets in 4-dimensional space. 

Let me finally comment on the ambiguity in defining the $Z_2$ 
gauge theory by projection. $Z_2$ projected theories which are 
constructed from gauge invariant variables and which do not involve 
gauge fixing before projection have been discussed in the literature 
for more than twenty years~\cite{mack,tom81}. These projection 
techniques preserve the value of the string tension~\cite{tom97} 
while the properties of the corresponding vortices do not have 
a physical interpretation in the continuum limit~\cite{fab99}. 
It was first noticed in~\cite{kurt} that the vortices which emerge 
from the construction outlined below (\ref{eq:1}) 
are physical objects (see also~\cite{eng98}).

\subsection{ Vortex properties } 

In the pioneering work~\cite{deb96}, it was obtained 
that the $Z_2$ projection of SU(2) gauge theory as discussed below 
(\ref{eq:1}) 
does not change the string tension. Figure \ref{fig:1} illustrates this 
fact using our results. At small quark anti-quark distances $r$, 
the Creutz ratios of the full SU(2) gauge theory show a $1/r^2$ 
behavior due to Coulomb interactions while they approach the 
string tension at large distances. The Creutz ratios of the $Z_2$ 
projected theory are (almost) constant. The Coulomb part is missing, 
whereas the value of the string tension is reproduced within statistical 
errors. By the projection $SU(2) \rightarrow Z_2$ along the lines 
described above the number of degrees of freedom is strongly reduced while 
the confining properties are preserved. Further numerical evidence 
that confinement is intrinsically related to the center vortices 
was accumulated in~\cite{deb98}. 

It was pointed out for the first time in~\cite{kurt,eng98}, and 
subsequently confirmed in~\cite{deb98}, that the vortices which 
arise from the $Z_2$ projected theory acquire physical relevance. 
In particular, it was observed that the density $\rho $ of vortices which 
pierce a hyperplane of space-time extrapolates to the continuum limit 
(see also figure \ref{fig:1}). Choosing the string tension $\sigma 
= (440 \, \hbox{MeV})^2 $ as reference scale, one finds 
$\rho \approx 3.5 \, / \hbox{fm}^{-2}$~\cite{eng98}. Moreover, 
a significant correlation between the vortices survive the continuum limit. 
This correlation amounts for an attraction up to a distance scale of 
$\approx 0.4 \, $fm~\cite{eng98}.

\section{ Vortices at finite temperatures } 

\begin{figure}[t]
\rule{5cm}{0.2mm}\hfill\rule{5cm}{0.2mm}
\centerline{
\hspace{-.6cm} 
\psfig{figure=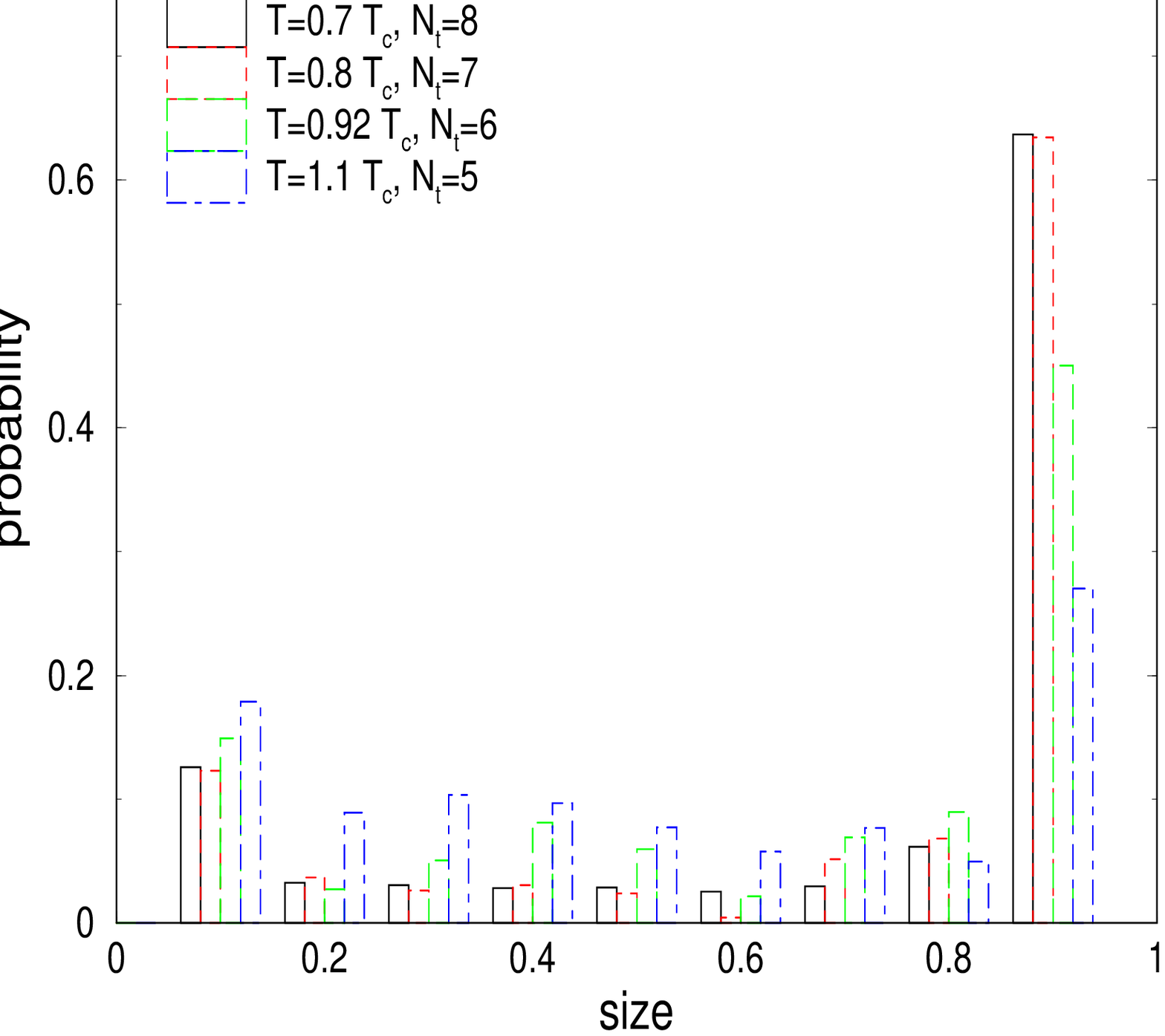,height=1.5in}
\hspace{1cm} 
\psfig{figure=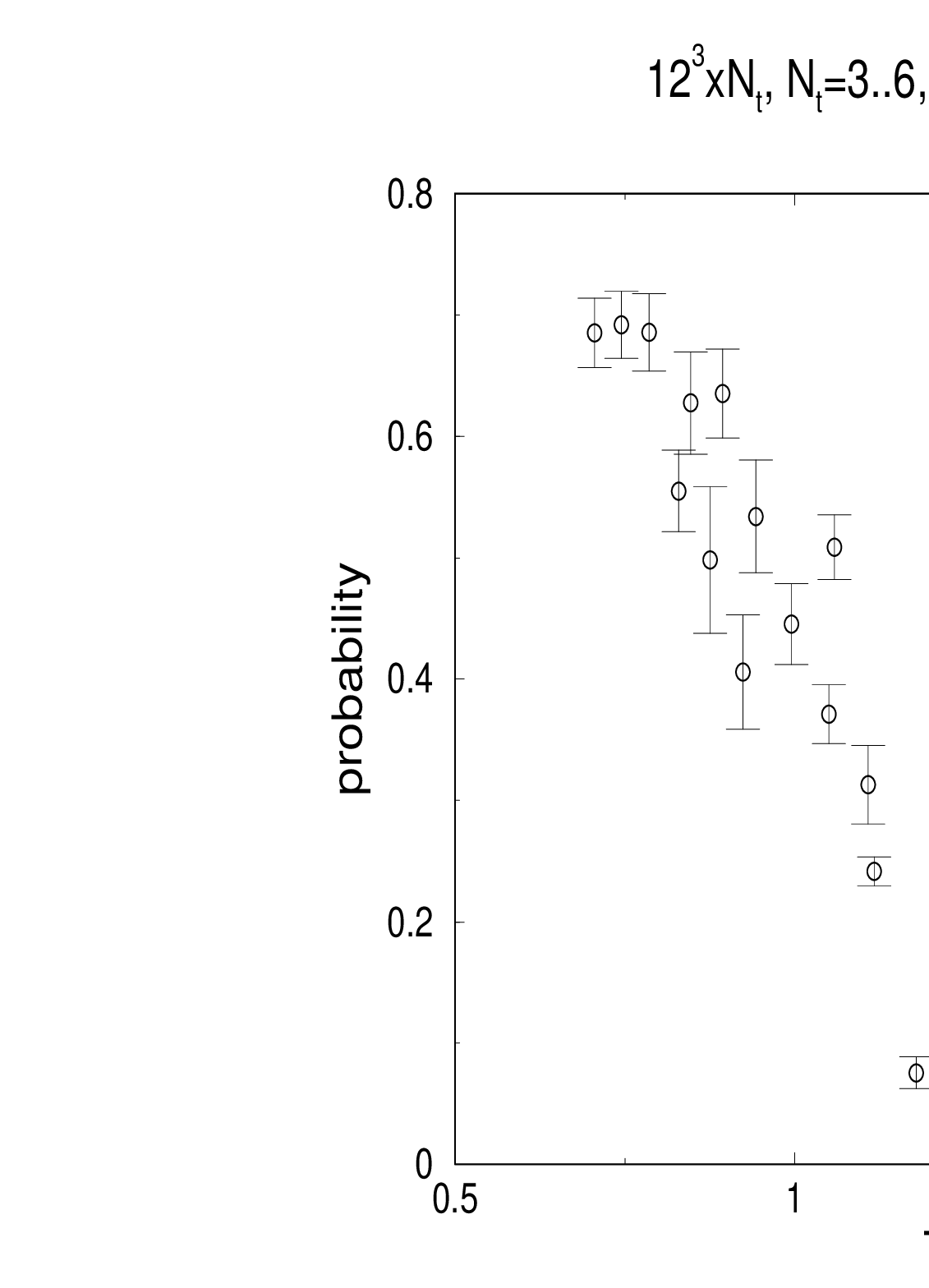,height=1.5in}
}
\rule{5cm}{0.2mm}\hfill\rule{5cm}{0.2mm}
\caption{ Cluster size distributions (left panel); percolation 
  probability (right panel).   \label{fig:2}}
\end{figure}
Finite temperatures are introduced in field theory by imposing 
periodic boundary conditions to Bose fields in time direction. The 
length of periodicity $L_t=1/T$ refers to the temperature $T$. 
It was argued in~\cite{app81} that 4-dimensional Yang-Mills theory 
at asymptotic high temperatures effectively behaves like the 
3-dimensional counter part the coupling constant of which is related 
to temperature\footnote{This mechanism of {\it dimensional reduction} 
was nicely confirmed by direct lattice calculations~\cite{bal93}.}. 
Since 3-dimensional Yang-Mills theory does confine quarks, a theory 
which is supposed to describe the deconfinement phase transition 
must cover the following phenomenon: above the critical temperature 
the string tension which is extracted from Polyakov lines must 
vanish, while correlations between Polyakov lines which are embedded in the 
spatial hyper cube signal a with temperature increasing ''spatial'' 
string tension.

\subsection{ Vortex polarization and pairing } 

In order to reveal a possible connection between vortex properties and 
the deconfinement phase transition at finite temperatures, we first 
showed that the $Z_2$ projected theory correctly describes the 
temperature dependence of string tension~\cite{eng99}. In particular, 
we find that the $Z_2$ theory correctly reproduces the critical temperature 
$T_c \approx 300 \, $MeV~\cite{eng99}. 

Keeping in mind that the expectation value of a Wilson loop is 
given in the vortex state by $\langle W \rangle = \sum _n (-1)^n \, P(n)$, 
where $P(n)$ is the probability that $n$ vortices pierce the 
minimal area of the loop, the vortex state might account for 
dimensional reduction if the vortices are polarized at high temperatures. 
If the vortices are aligned along the time axis direction by 
temperature effects, they would not pierce Wilson areas which are spanned 
by the time axis and one spatial direction. This would account for 
a vanishing string tension. On the other hand, there would be plenty 
of vortices randomly piercing spatial Wilson loops thus supporting a 
non-vanishing spatial string tension. In order to test this idea of the 
deconfinement phase transition, we investigated the density of vortices 
piercing time-like and space-like oriented planes. We found that the 
spatial vortex density indeed increases at large temperatures~\cite{la99} 
and that the corresponding spatial string tension is indeed compatible with 
ideas of dimensional reduction~\cite{eng99}. However, the time-like vortex 
density has only dropped by a factor of two at twice the critical 
temperature~\cite{la99} implying that the sharp decrease of the string 
tension at the critical temperature is not reflected by the (time-like) 
vortex density.

\subsection{ Vortex percolation } 

All the numerical results which have been obtained so far suggest 
an intimate relation of confinement and vortex 
percolation~\cite{la99,la99c,eng99}: 
for $T<T_c$ the vortices percolate and generically form a cluster 
which fills the (lattice) universe. For $T>T_c$ the vortices cease 
to percolate. The large vortex cluster decays into smaller ones. 
If the average cluster size is smaller than the length scale of the Wilson 
loop, the vortices which are located off the Wilson loop 
boundaries must pierce the 
loop an even number of times. Hence, they do not contribute to the 
Wilson expectation value. Only vortices close to the periphery of 
the loop provide non-trivial contributions, and the Wilson loop 
expectation value exhibits a perimeter law signaling 
deconfinement. In order to underline this picture of the deconfinement 
phase transition, we have calculated the probability $p(x)$ that 
a vortex link is part of vortex cluster of size $x$. Thereby, $x=1$ 
corresponds to the maximum size of the lattice universe. Figure 
\ref{fig:2} shows $p(x)$ for several values of the 
temperature~\cite{eng99}. A clear signal for vortex percolation is found 
for small temperatures whereas for $T>T_c$ small vortex cluster sizes 
dominate~\cite{eng99}. Also shown in figure \ref{fig:2} is the percolation 
probability, i.e. $p( x>0.9)$, as function of temperature. It sharply 
decreases at $T=T_c$. 

\section*{Acknowledgments}
It is a pleasure to thank my collaborators M.~Engelhardt, H.~Reinhardt, 
O.~Tennert. I am indebted to the organizers, in particular David Blaschke, 
for giving me the possibility to participate in this interesting 
workshop.


\section*{References}
\begin{thebibliography}{99}
\bibitem{tho81} G.~'t~Hooft, \Journal{\NPB}{190}{455}{1981}.  
\bibitem{bali} K.~Schilling, G.~S.~Bali, C.~Schlichter,  
  Lattice 98, hep-lat/9809039. 
\bibitem{deb96} L.~Del Debbio, M.~Faber, J.~Greensite and 
  \v{S}.~Olejn{\'\i}k, \Journal{{\em Nucl. Phys. Proc. Suppl. }}{53}{141}
  {1997}. 
\bibitem{deb97} L.~Del Debbio, M.~Faber, J.~Greensite and
  \v{S}.~Olejn{\'\i}k, \Journal{\PRD}{55}{2298}{1997}. 
\bibitem{mack} G.~Mack and V.~B.~Petkova, \Journal{ {\rm Ann. Phys. (NY)}}
  {123}{442}{1979}; \Journal{{\em Ann. Phys. (NY)}}{125}{117}{1980}; \\
  G.~Mack, \Journal{\PRL}{45}{1378}{1980}; \\ 
  G.~Mack and E.~Pietarinen, \Journal{\NPB}{205}{141}{1982}. 
\bibitem{tom81} E.~T.~Tomboulis, \Journal{\PRD}{23}{2371}{1981}; 
  \Journal{\PLB}{303}{103}{1993}. 
\bibitem{tom97} T.~G.~Kov\'acs and E.~T.~Tomboulis,
  \Journal{\PRD}{57}{4054}{1998}; \\ 
  J.~Ambj{\o}rn and J.~Greensite, \Journal{{\em JHEP}}{9805}{004}{1998}. 
\bibitem{fab99} M.~Faber, J.~Greensite and \v{S}.~Olejn{\'\i}k,
  JHEP {\bf 9901} (1999) 008.
\bibitem{deb98} L.~Del Debbio, M.~Faber, J.~Giedt, J.~Greensite and 
  S.~Olejnik, \Journal{\PRD}{58}{094501}{1998}. 
\bibitem{kurt} K.~Langfeld, H.~Reinhardt and O.~Tennert,
  \Journal{\PLB}{419}{317}{1998}. 
\bibitem{eng98} M.~Engelhardt, K.~Langfeld, H.~Reinhardt and O.~Tennert,
  \Journal{\PLB}{431}{141}{1998}. 
\bibitem{app81} T.~Appelquist and R.~D.~Pisarski, 
  \Journal{\PRD}{23}{2305}{1981}. 
\bibitem{bal93} G.~S.~Bali, J.~Fingberg, U.~M.~Heller, F.~Karsch
  and K.~Schilling, \Journal{\PRL}{71}{3059}{1993}. 
\bibitem{eng99} M.~Engelhardt, K.~Langfeld, H.~Reinhardt and O.~Tennert,
  hep-lat/9904004.
\bibitem{la99} K.~ Langfeld, O.~Tennert, M.~Engelhardt and H.~Reinhardt, 
  hep-lat/9805002, in press by {\PLB}. 
\bibitem{la99c} Kurt Langfeld, Presented at 4th Workshop on Quantum 
   Chromodynamics, Paris, France, 1-6 Jun 1998, hep-lat/9809003. 

\end{thebibliography}
\end{document}